%% file: main.tex
\documentclass[conference]{IEEEtran}
\usepackage{balance}

\usepackage{tikz}
\usetikzlibrary{arrows.meta,positioning,fit,shapes,backgrounds}
\usepackage{listings}
\usepackage{lscape}
\usepackage{minted}
\usepackage{dirtree}

\usepackage[inkscapeformat=png]{svg}

\usepackage{enumitem}
\usepackage{graphicx}
\usepackage{multicol, latexsym}
\usepackage{blindtext}
\usepackage[skip=5pt,font=small]{caption}
\usepackage{longtable}
\usepackage{csquotes}
\usepackage{amsfonts}
\usepackage{amsmath}
\usepackage{amsthm}
\usepackage{amssymb}
\usepackage{algorithm}
\usepackage{tabularx}
\usepackage{capt-of,lipsum} 
\usepackage{subcaption}

\usepackage{listings}
\usepackage{caption}
\usepackage{comment}
\usepackage{diagbox}
\usepackage{pdflscape}
\usepackage{booktabs}
\usepackage{makecell}
\usepackage{multirow}
\usepackage{fourier} 
\usepackage{array}

\usepackage{siunitx}
\usepackage{mathtools, nccmath}

\usepackage{siunitx}
\DeclareSIUnit[number-unit-product = { }] \dBm{dBm}

\usepackage{algpseudocode}
\usepackage{algorithm}

\usepackage{listing-styles}
\usepackage{wasysym}

\algnewcommand\algorithmicforeach{\textbf{for each}}
\algdef{S}[FOR]{ForEach}[1]{\algorithmicforeach\ #1\ \algorithmicdo}

\usepackage[hidelinks]{hyperref}

\definecolor{myblue}{rgb}{0.09,0.20,0.34}
\definecolor{mygreen}{rgb}{0,0.6,0}
\definecolor{mygray}{rgb}{0.98,0.98,0.98}
\definecolor{myorange}{rgb}{0.92,0.49,0.34}
\definecolor{mywhite}{rgb}{1.0,1.0,1.0}

\definecolor{NMR}{RGB}{255,255,86}
\definecolor{MRA}{RGB}{255,231,27}
\definecolor{MRD}{RGB}{178,178,178}
\definecolor{MRIP}{RGB}{188,172,0}
\definecolor{MRS}{RGB}{161,207,106}
\definecolor{NA}{RGB}{228,60,52}
\definecolor{AI}{RGB}{255,255,86}
\definecolor{RMR}{RGB}{162,4,21}
\definecolor{RMD}{RGB}{178,178,178}
\definecolor{RMA}{RGB}{255,231,27}
\definecolor{RMIP}{RGB}{188,172,0}
\definecolor{RMS}{RGB}{161,207,106}
\definecolor{LIGHT_GREY}{RGB}{240,240,240}
\definecolor{lightgray}{rgb}{.9,.9,.9}
\definecolor{darkgray}{rgb}{.4,.4,.4}
\definecolor{purple}{rgb}{0.65, 0.12, 0.82}
\definecolor{darkgreen}{RGB}{30, 142, 20}

\newcommand{\ie}{\textit{i}.\textit{e}.,\ }
\newcommand{\eg}{\textit{e}.\textit{g}.,\ }
\newcommand{\cf}{\textit{c}\textit{f}.\ }

\newcommand\1{\textit{(i)}}
\newcommand\2{\textit{(ii)}}

\newcommand\4{\textit{(iv)}}

\newsavebox{\mybox}

\usepackage{pifont}

\usepackage{censor}
\usepackage[scaled]{helvet}
\usepackage[T1]{fontenc}
\usepackage{helvet}

\usepackage{tabularx,booktabs,makecell,array}
\newcolumntype{Y}{>{\centering\arraybackslash}p{0.06\textwidth}} 

\begin{document}





\title{Distributed Pulse-Wave Simulator\\for DDoS Dataset Generation}


\DeclareRobustCommand*{\IEEEauthorrefmark}[1]{%
  \raisebox{0pt}[0pt][0pt]{\textsuperscript{\footnotesize #1}}%
}

\author{\IEEEauthorblockN{ 
Karim Khamaisi\IEEEauthorrefmark{2},
Pascal Kiechl\IEEEauthorrefmark{1},
Katharina Müller\IEEEauthorrefmark{1},
Burkhard Stiller\IEEEauthorrefmark{1},
Bruno Rodrigues\IEEEauthorrefmark{2}
}


\IEEEauthorblockA{\IEEEauthorrefmark{1}Communication Systems Group CSG, Department of Informatics IfI, 
University of Zurich UZH, Switzerland\\
}

\IEEEauthorblockA{\IEEEauthorrefmark{2}Embedded Sensing Group ESG, School of Computer Science SCS, 
University of St. Gallen HSG, Switzerland\\
} 
E-mail:pascal.kiechl@uzh.ch, [stiller|mueller]@ifi.uzh.ch, [bruno.rodrigues|karim.khamaisi]@unisg.ch
}


\newtheoremstyle{mydef}
{\topsep}{\topsep}%
{}{}%
{\bfseries}{}
{\newline}
{%
  \rule{\linewidth}{0.4pt}\\*%
  \thmname{#1}~\thmnumber{#2}\thmnote{\ -\ #3}.\\*[-1.5ex]%
  \rule{\linewidth}{0.4pt}}%
\theoremstyle{mydef}
\newtheorem{definition}{Definition}
\newtheorem{protocol}{Step}

            


\maketitle



\begin{abstract}
Pulse-wave Distributed Denial-of-Service (DDoS) attacks generate short, synchronized bursts of traffic that circumvent pattern-based detection and quickly exhaust traditional defense systems. This transient and spatially distributed behavior makes analysis extremely challenging, as no public datasets capture how such attacks evolve across multiple network domains. Since each domain observes only a partial viewpoint of the attack, a correlated, multi-vantage view is essential for comprehensive analysis, early detection, and attribution.

This paper presents \emph{DPWS}, an open-source simulator for generating distributed pulse-wave DDoS datasets. DPWS models multi-AS topologies and produces synchronized packet captures at multiple autonomous systems, showing the distributed structure of coordinated bursts. It enables fine-grained control of traffic parameters through a lightweight YAML interface. DPWS reproduces pulse-wave dynamics across multiple vantage points, exhibits natural fingerprint variability at equal aggregate rates, and scales with MPI in ns-3, providing a reproducible basis for studying pulse-wave behaviour and benchmarking distributed DDoS defenses, while sharing practical insights on ns-3 scalability and synchronization gained during development.
\end{abstract}





%

\IEEEpeerreviewmaketitle

\pagestyle{plain} 


\input{sections/1_introduction}
\input{sections/2_fundamentals}
\input{sections/3_design}
\input{sections/4_experiments}

\input{sections/5_final_considerations}







\bibliographystyle{IEEETranS}
\bibliography{bib/main.bib}


\end{document}

%% file: sections/1_introduction.tex
\section{Introduction} \label{chp:introduction}

Distributed Denial-of-Service (DDoS) are among the oldest and most persistent threats in the history of the Internet. Despite decades of research and the deployment of large-scale mitigation infrastructures, they remain difficult to detect and counter effectively. For example, in May 2025, CloudFlare \cite{cloudflareDDoSFamous} reported the largest DDoS attack ever seen, peaking 7.3 Tbps (Terabits per second) of traffic. The attack exploited multiple vectors including UDP floods, NTP reflection, and traffic from known botnets (\eg Mirai). The provider reported that the attack was successfully mitigated via a coordinated operation of its globally distributed protection network.


Pulse-wave attacks have recently been studied, providing a deeper understanding of their characteristics (also quantified based on real-world datasets \cite{kopp2025ddos}), and mitigation alternatives exploring programmable switches and aggregate-based congestion control \cite{alcoz2022aggregate-based}. Still, no single network domain can observe the full spatio-temporal evolution of such coordinated attacks, and a distributed view across multiple vantage points is essential to correlate partial observations and enable timely mitigation. For example, as CloudFlare leveraged its network to mitigate the record 7.3 Tbps attack.


Data used to design, train, and evaluate defenses rarely present this cooperative reality. Widely used resources (\ie datasets) such as CIC-DDoS2019 \cite{cicddos2019}, UNSW-NB15 \cite{unsw2015nb15}, and long standing traces like CAIDA 2007 \cite{caida2007ddos} or MAWILab \cite{mawilab} are still very valuable. However, they typically lack support for multiple ISP vantage points, inter domain signaling and enforcement artifacts such as DOTS, FlowSpec, and RTBH, and operator oriented labels that track playbooks and mitigation life cycles. Thus, models tuned in a single environment often fail to generalize under domain shift during real incidents \cite{kopp2025ddos}. The scenario can be even more complex as many reflection and amplification attacks still rely on spoofed sources, which persists due to the uneven deployment of source address validation. These cases also reinforces the need for mitigation closer to the source and for distributed detection (as in our previous work \cite{brunner2023deciphering}) that looks beyond the victim.

\begin{table*}[t]
\centering
\footnotesize
\setlength{\tabcolsep}{4pt}
\renewcommand{\arraystretch}{1.1}
\begin{tabularx}{\textwidth}{@{}p{0.16\textwidth} p{0.20\textwidth} *{5}{>{\centering\arraybackslash}p{0.06\textwidth}} X@{}}
\toprule
\textbf{Class / Work} & \textbf{Representative refs} &
\textbf{Multi ISP} & \textbf{DOTS} &
\textbf{FlowSpec or RTBH} &
\textbf{Operator labels} & \textbf{Drills} & \textbf{Notes} \\
\midrule
Public traces & CAIDA~2007, MAWILab~\cite{caida2007ddos,mawilab} & \Circle & \Circle & \Circle & \Circle & \Circle & Single vantage only. \\
ML datasets & CIC-DDoS2019, UNSW-NB15~\cite{cicddos2019,unsw2015nb15} & \Circle & \Circle & \Circle & \Circle & \Circle & Labeled flows, single domain. \\
Traffic generators & D-ITG, ApacheBench~\cite{botta2012tool,apacheBench} & \Circle & \Circle & \Circle & \Circle & \Circle & Stress testing tools. \\
Real testbeds & PlanetLab, GENI~\cite{peterson2006experiences,elliott2008geni} & \LEFTcircle & \Circle & \Circle & \Circle & \Circle & Real stacks, not standards aware. \\
NDT frameworks & ITU-T~Y.3090, IRTF~NMRG~\cite{itutY3090,nmrgnetworkdtwin} & \LEFTcircle & \Circle & \Circle & \Circle & \LEFTcircle & Generic twins, not DDoS specific. \\
\textbf{DPWS} & this work & \CIRCLE & \LEFTcircle & \LEFTcircle & \CIRCLE & \CIRCLE & Multi-domain twin with pcaps; architecture designed for DOTS and FlowSpec/RTBH integration. \\
\bottomrule
\end{tabularx}
\caption{Comparison of related efforts. (Legend: \CIRCLE: provided, \Circle: not provided, \LEFTcircle: partially provided).}
\label{tab:related}
\end{table*}


This paper presents \textit{Distributed Pulse‑wave DDoS Simulator} (DPWS) \cite{dpws-code}, a system that creates "twin" scenarios in the \emph{ns-3} discrete event simulator \cite{ns3}. DPWS produces time synchronized packet captures and flow records per domain \cite{rfc8811}, and is designed to include mitigation artifacts from FlowSpec and RTBH \cite{rfc5635}. It combines data plane traces with the corresponding control plane actions. Attack strategies covers volumetric floods and state exhaustion, and reflection/amplification and application layer HTTP floods. Recording occurs at backbone and border routers and at the target. 

DPWS makes the following contributions:
\begin{itemize}
  \item \textbf{Traffic models guided by public DDoS traces} (\eg CAIDA \cite{caida2007ddos}, MAWI \cite{mawilab}) that reproduce benign flows and ICMP or SYN flood dynamics, with clear paths to reflection and application layer floods.
  \item \textbf{Traffic models based on public traces} that reproduce benign flows and ICMP or SYN flood dynamics, with clear paths to reflection and application layer floods.
  \item An \textbf{open and reproducible workflow} configurable in YAML with a  command line interface, exporting \textit{pcap} datasets for drills and evaluation.
\end{itemize}

The paper is organized as follows. Section~\ref{sec:related-work} presents related work, Section~\ref{sctn::design} presents DPWS design. Section~\ref{sec:evaluation} reports the evaluation and discussion, and Section~\ref{sec:final_considerations} concludes the paper.


%% file: sections/2_fundamentals.tex
\section{Background} \label{sec:related-work}
A pulse-wave is a relatively recent form of DDoS attack characterized by short, high-rate traffic bursts that may alternate between targets or protocols~\cite{kopp2025ddos}. Unlike conventional DDoS attacks, which gradually ramp up and maintain a steady rate~\cite{alcoz2022aggregate-based}, pulse-wave attacks shows a rapid on–off pattern that can overwhelm mitigation systems relying on delayed reaction.

\begin{figure}[H]
\centering
   \begin{subfigure}{\columnwidth} \centering
    \includegraphics[scale=0.32, trim={0cm 10cm 0 0},clip]{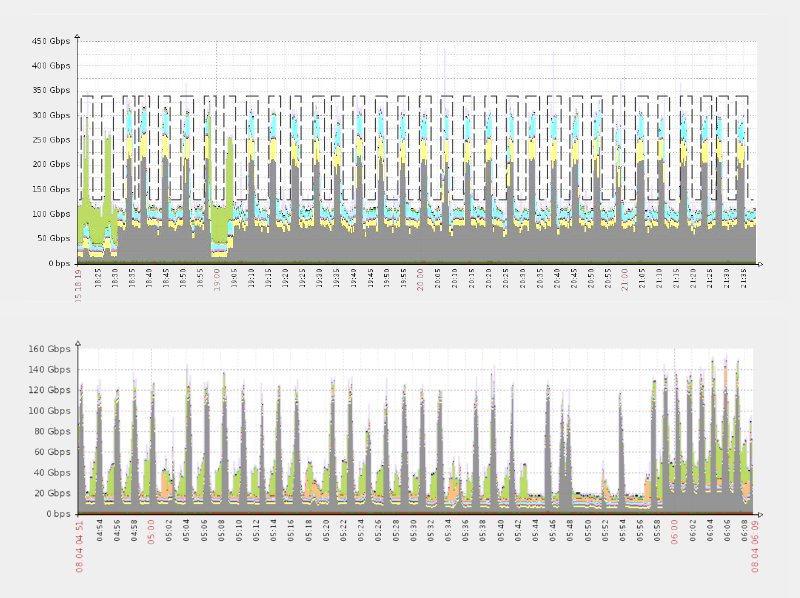}
    \caption{Pulse-Wave DDoS Attack}
    \label{fig::bg::pw_vs_normal::pw}
   \end{subfigure}
   \\
   \begin{subfigure}{\columnwidth} \centering
     \includegraphics[scale=0.32, trim={0cm 0cm 0 8cm},clip]{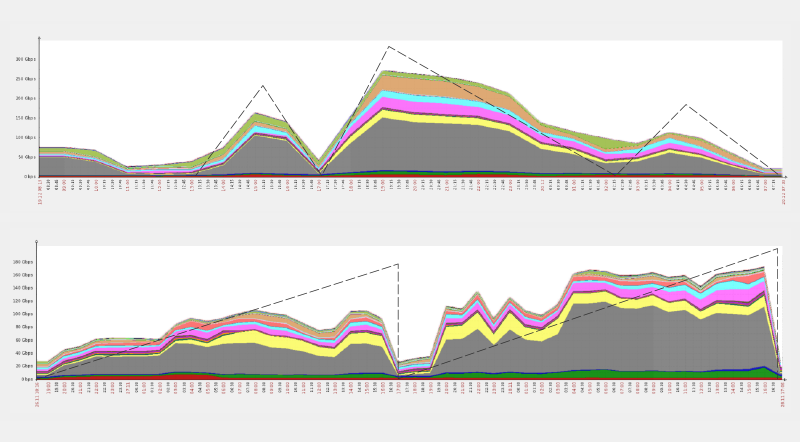}
    \caption{Traditional DDoS Attack}
    \label{fig::bg::pw_vs_normal::traditional}
   \end{subfigure}
\caption{Comparison of Pulse-Wave and traditional DDoS attack Mechanisms~\cite{imperva2017pulsewave}} \label{fig::bg::pw_vs_normal}
\end{figure}

Figure~\ref{fig::bg::pw_vs_normal::pw} shows the characteristic burst pattern of a pulse-wave attack, while Figure~\ref{fig::bg::pw_vs_normal::traditional} depicts a conventional DDoS with gradual buildup and sustained load. The contrast shows how pulse-wave attacks exploit reaction delays in hybrid mitigation systems making them extremely difficult to be detected and mitigated without a global overview.

\section{Related Work}

Table~\ref{tab:related} shows representative work along main criteria to DDoS defense, such as multi-ISP vantage points, inter-domain signaling (\eg DOTS), enforcement artifacts (FlowSpec and RTBH), operator-centric labels, and support for repeatable DDoS drills. 

\noindent \textbf{Attack tools and traffic generators.}
Foundational surveys synthesize the evolution from early tools to modern botnet-driven capabilities and highlight their value for stress testing IDS and scrubbing stacks. These tools generate controllable load and attack mixes, yet lack distributed vantage points and do not emit control-plane or mitigation artifacts needed to study coordinated defense \cite{behal2017characterization}.

\noindent \textbf{Existing datasets.}
Historic traces such as CAIDA 2007 remain pedagogical references \cite{caida2007ddos}, while newer efforts like CIC-DDoS2019 \cite{cicddos2019} add labeled benign and attack flows. Despite their utility for training anomaly-based IDS, most datasets expose only a single capture point and often sanitize headers, which removes inter-domain context and operational actions that matter for end-to-end defense evaluation \cite{sharafaldin2019developing}.

\noindent \textbf{Dataset generation approaches.}
Three families dominate. Simulation frameworks (\eg NS-3, OMNeT++ \cite{ns3, omnet++}) enable repeatable studies and large parameter sweeps, but struggle to replay extreme floods with protocol fidelity. Emulation adds realism with real stacks over virtual links, though scale quickly becomes a constraint. Planet-scale testbeds offer high realism but are costly to operate, risky for live malware, and hard to reconfigure. These show important facets of the problem but rarely pair distributed visibility with inter-domain mitigation context \cite{peterson2006experiences}.

\noindent \textbf{Inter-domain mitigation standards.}
Operational mitigation increasingly depends on cross-domain signaling and rapid enforcement. DOTS \cite{rfc8811} defines an authenticated signal channel with telemetry and status exchange. On the enforcement path, BGP FlowSpec accelerates filter dissemination \cite{rfc8955} and RTBH \cite{rfc5635} enables targeted traffic shedding. Public datasets almost never include these artifacts, which limits faithful evaluation of coordinated workflows \cite{itutY3090}.

Prior work fragments into single-domain traces, synthetic datasets tuned for ML, flexible generators, or high-realism testbeds. None combine multi-ISP visibility with inter-domain signaling and mitigation. DPWS presents a multi-domain simulator that creates synchronized \textit{pcaps} and flows together with DOTS telemetry, FlowSpec and RTBH enforcement logs, enabling repeatable drills and cross-domain evaluation.

%% file: sections/3_design.tex
\section{DPWS Overview}
\label{sctn::design}

\begin{figure}[h]
    \centering
    \includegraphics[width=1\columnwidth]{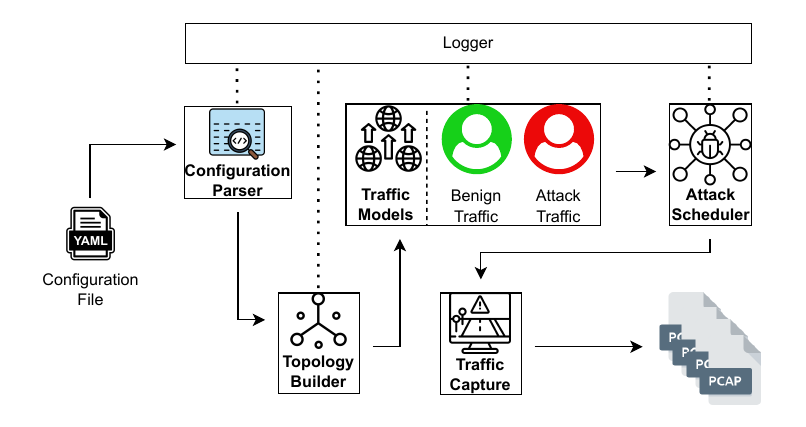}
    \caption{System Architecture}
    \label{fig::design::architecture}
\end{figure}

The \textit{Distributed Pulse‑Wave Simulator} (DPWS) models multi-domain DDoS scenarios in ns-3 to produce synchronized packet traces from several network vantage points. Figure~\ref{fig::design::architecture} outlines its architecture, which comprises six modular components: a configuration parser, topology builder, traffic models, attack scheduler, capture subsystem, and logger.

\begin{itemize}
    \item \textbf{Configuration Parser}: loads YAML configuration, validates topology and fills defaults.
    \item \textbf{Topology Builder}: creates an Internet-exchange-like Central Network (CN) with attached Autonomous Systems (AS).
    \item \textbf{Traffic Models}: bind behaviors to nodes: benign HTTP flows and attacker traffic based on ICMP, UDP, or TCP floods.
    \item \textbf{Attack Scheduler}: coordinates pulse timing and vector switching among attacker nodes.
    \item \textbf{Capture Subsystem}: records packet traces at all CN interfaces, producing vantage-specific PCAPs for distributed analysis.
    \item \textbf{Logger}: provides unified runtime feedback and complete configuration output for reproducibility.
\end{itemize}

DPWS creates a set of Autonomous Systems interconnected through a Central Network that acts as an Internet-exchange fabric. Each AS hosts client and server nodes assuming attacker, benign, or non-target roles, with gateways linking to CN nodes where traffic captures occur.

\subsection{Topology Builder}
\label{sctn::design::topology}

The topology builder constructs an Internet-exchange-like \emph{Central Network} (CN) and attaches multiple \emph{Autonomous Systems} (AS) through gateway nodes. 
Each AS hosts clients and servers configured with specific roles, while CN links represent high-capacity inter-domain connections.

\begin{figure}[h]
    \centering
    \includegraphics[width=0.96\columnwidth,trim={0 0 0 0},clip]{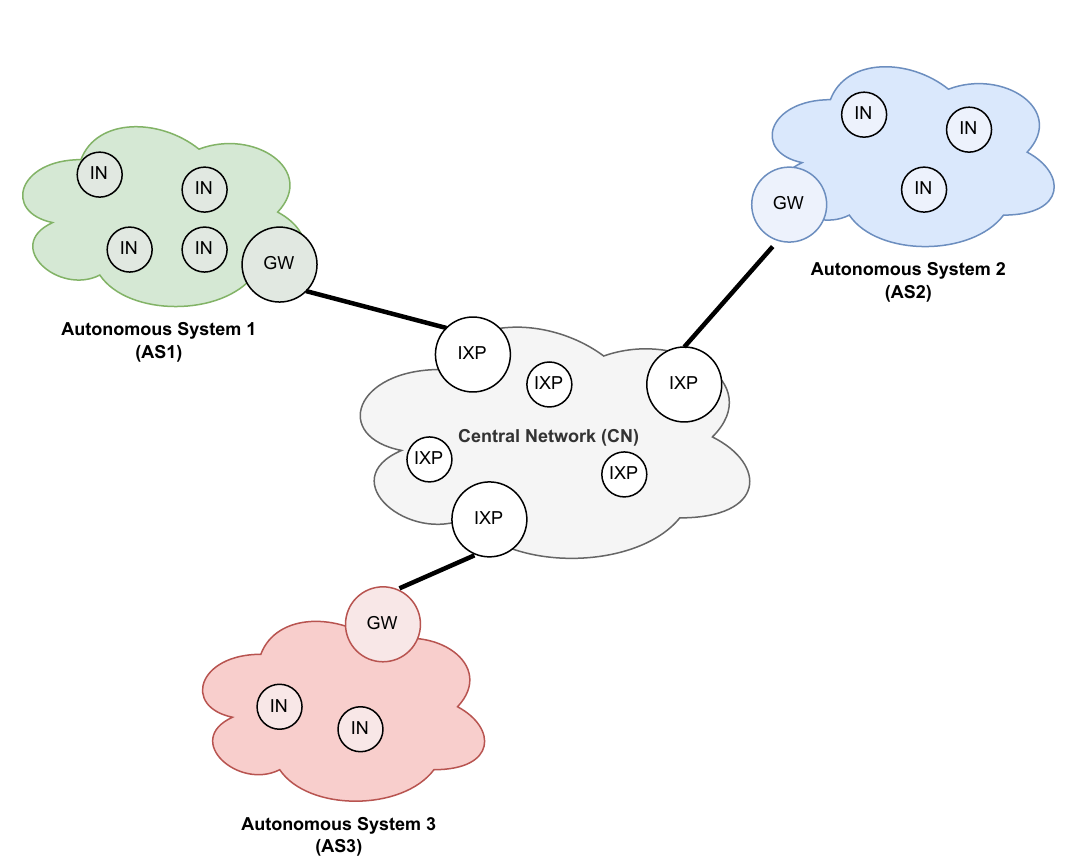}
    \caption{Topology construction workflow: randomized CN mesh (top) with degree of redundancy~$\rho$, AS attachments via gateways, and role assignment within each AS.}
    \label{fig::design::topology}
\end{figure}

The CN is generated as a partial mesh whose redundancy factor~$\rho\!\in[0,1]$ defines the number of superfluous links beyond the minimal spanning topology.  
Each AS connects through a single gateway to a CN node, and its internal links use point-to-point channels in dedicated subnets to avoid routing ambiguity.  
Node roles—attacker, benign, target, or non-target—are then instantiated and bound to corresponding traffic models.

\subsection{Traffic Generation Model}
\label{sctn::design::traffic-model}

To create the traffic generation strategy, DPWS instantiates an \emph{OnOffRetargetApplication} for each attacker and attack vector. Each instance alternates between \textit{on} and \textit{off} phases, modulated by per-vector parameters for burst duration $b_v$, switch interval $s_v$, and nominal rate $\bar r_{i,v}$. The sequence of targets $\mathcal{T}$ and vectors $\mathcal{V}$ defines a repeating cycle of length 
$C=\sum_{v\in\mathcal{V}}\!\big(|\mathcal{T}|\,b_v+(|\mathcal{T}|{-}1)\,s_v\big)$.
During each burst, packets of size $S_{i,v}$ are sent at rate
$r_{i,v}(t)=\bar r_{i,v}(1+\varepsilon_{i,v}(t))$, 
where $\varepsilon_{i,v}(t)\!\sim\!\mathcal{U}[-\delta_i,\delta_i]$ introduces per-packet jitter.

\begin{algorithm}[H]
\caption{Pulse-Wave Traffic Generation (per attacker $i$)}
\label{alg:traffic}
\begin{algorithmic}[1]
\For{each attack vector $v \in \mathcal{V}$}
    \State set start offset $o_v$ to serialize vectors
    \For{each target $k \in \mathcal{T}$}
        \State wait until $t = o_v + (k{-}1)(b_v + s_v)$
        \State set remote $\leftarrow$ target($k$)
        \State turn \texttt{ON} for duration $b_v$
            \While{time $< b_v$}
                \State draw $\varepsilon \sim \mathcal{U}[-\delta_i,\delta_i]$
                \State send packet of size $S_{i,v}$ after delay 
                      $\Delta = S_{i,v}/[\bar r_{i,v}(1+\varepsilon)]$
            \EndWhile
        \State turn \texttt{OFF} for duration $s_v$
    \EndFor
\EndFor
\end{algorithmic}
\end{algorithm}

Each attacker produces a periodic pulse-train whose amplitude (rate) and duty cycle (burst versus switch duration) can differ per vector. Summing rates across attackers and paths yields the instantaneous load at any link $e$ as
\[
R_e(t)=\sum_{i,v,k} r_{i,v}(t)\,u_{v,k}(t)\,\mathbf{1}\{\text{path}(i\!\to\!k)\ni e\},
\]
where $u_{v,k}(t)\!\in\!\{0,1\}$ indicates whether vector~$v$ targets~$k$ at~$t$. This formulation mirrors the implementation logic in ns-3 while remaining reproducible and parameter-driven.

\subsection{Attack Scheduling}
\label{sctn::design::attack-scheduling}

The scheduler computes a single periodic timetable that all attacker applications follow. Each attacker creates one \texttt{OnOffRetargetApplication} per vector. The \texttt{AttackScheduleHelper} assigns a start offset $o_v$ per vector, burst and switch intervals $(b_v,s_v)$ per target, and the instants to call \texttt{SetRemote()} to retarget the application. The cycle length is
\[
C=\sum_{v\in\mathcal{V}}\Big(|\mathcal{T}|\,b_v+(|\mathcal{T}|-1)\,s_v\Big).
\]

\begin{figure}[h]
    \centering
    \includegraphics[width=1\columnwidth]{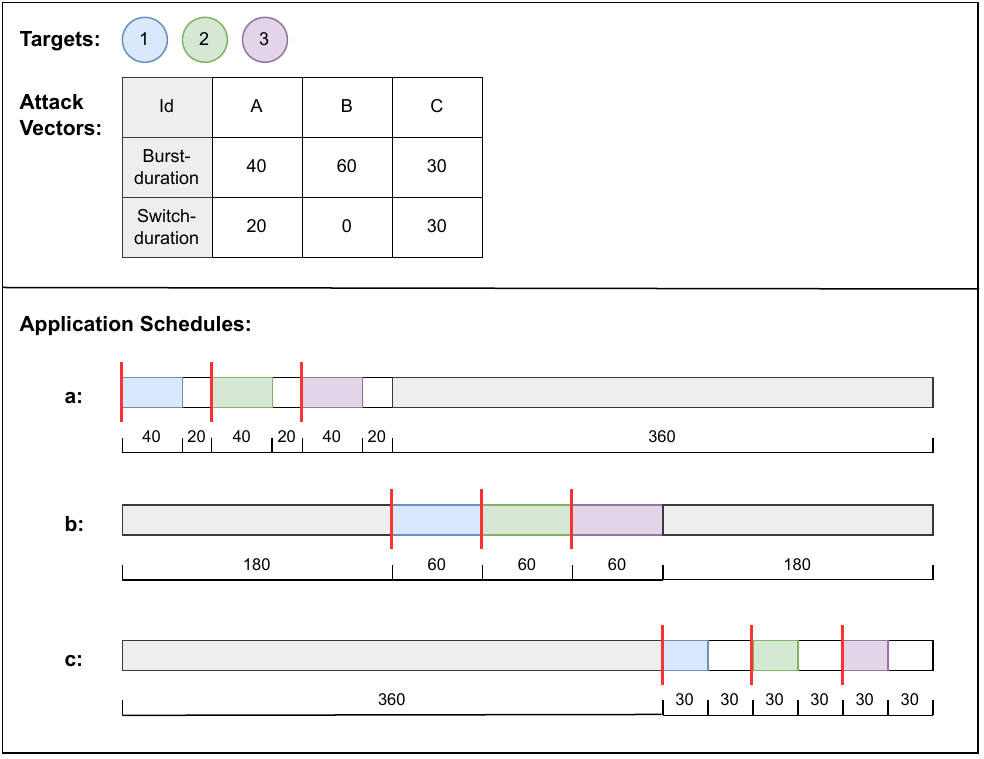}
    \caption{Example of attack scheduling vectors (duration in seconds).}
    \label{fig::impl::atk_schedule}
\end{figure}

Each vector \(v\) produces \(|\mathcal{T}|\) bursts of duration \(b_v\) and \(|\mathcal{T}|{-}1\) switch intervals \(s_v\). The total cycle length \(C\) therefore represents the sum of all active and switching phases across vectors, defining when the full pulse sequence repeats. Figure~\ref{fig::impl::atk_schedule} illustrates a complete scheduling cycle with three attack vectors and three targets. 

Each vector operates sequentially, generating short bursts (\textit{on-states}) separated by configurable switch intervals during which targets are retargeted. 
Colored blocks denote active bursts, white gaps indicate target switching or idle phases, and red markers represent \texttt{SetRemote()} events. 
Offsets between vectors ensure that only one attack pattern is active at a time, while long idle periods allow other vectors to complete their rounds before the cycle repeats. This coordination achieves realistic, alternating pulse-wave traffic across multiple attackers and targets.

\subsection{Capture and Logging Subsystem}
\label{sctn::design::capture}

The capture and logging components provide distributed observability and reproducibility across simulation runs. 
Packet capture is enabled on every interface of the CN, ensuring that traffic is recorded from multiple inter-domain vantage points. 
Each capture file is named according to the link direction, following the pattern  \texttt{\{prefix\}\_\_FromID-to-ToID\_\_\{suffix\}.pcap}, where \texttt{prefix} identifies the simulation scenario and \texttt{FromID}/\texttt{ToID} denote the connected CN or AS nodes. This naming scheme allows quick correlation of traces and facilitates comparative analysis across domains. The logger records configuration parameters, timestamps, and simulation events into a unified log, providing a full reproduction trail for each experiment. 

%% file: sections/4_experiments.tex
\section{Experimental Evaluation}\label{sec:evaluation}

This section evaluates DPWS across three aspects: distributed observability, attack diversity, and system scalability.


\subsection{Distributed Perspective}

\begin{figure}[h]
  \centering
  \includegraphics[width=\columnwidth]{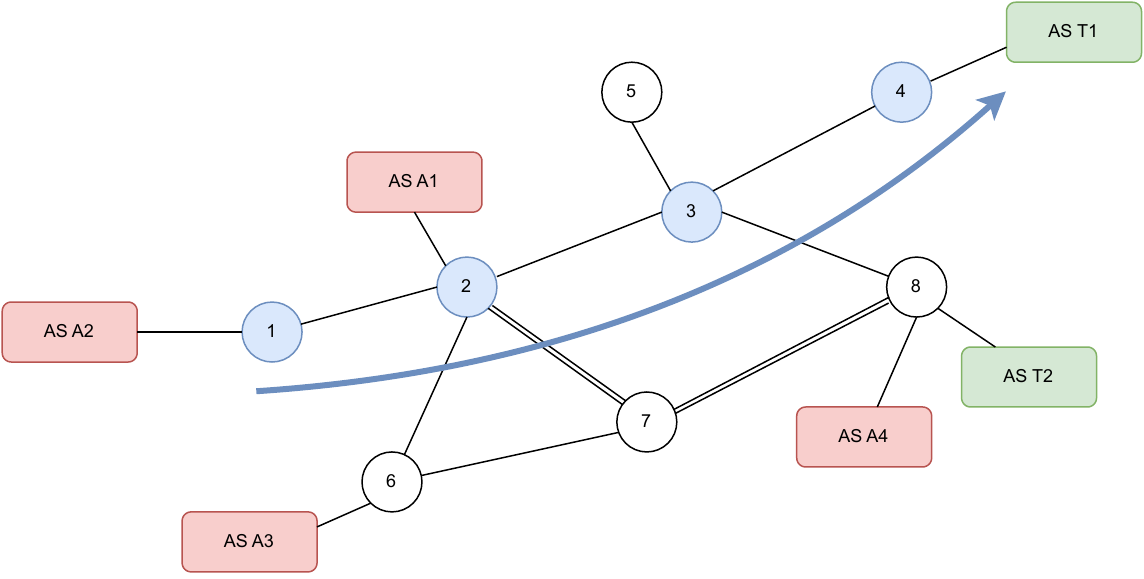}
  \caption{DIST topology and capture points. Observations diverge along the path as target selection and routing diversity manifest.}
  \label{fig::eval::dist::topology}
\end{figure}

The \textit{DIST} (Distributed Perspective) scenario illustrates how different vantage points observe the same attack. It includes eight CN nodes interconnecting four ASs hosting twelve attackers and two targets. Figure~\ref{fig::eval::dist::topology} shows the topology and capture points, while Table~\ref{tab::eval::dist} summarizes the configuration. A single CN path (in blue) is analyzed hop by hop to show how traffic aggregates near the source and splits toward distinct targets as routing and AS contributions diverge.

\begin{table}[h]
\caption{DIST scenario configuration}
\label{tab::eval::dist}
\centering
\begin{tabular}{l|c|c|c}
\toprule
\textbf{Scenario} & \textbf{CN nodes} & \textbf{Attackers} & \textbf{Targets} \\
\midrule
DIST & 8 & 12 (across 4 ASs) & 2 (in 2 ASs) \\
\bottomrule
\end{tabular}
\end{table}

Hop-wise captures along the selected CN path reveal where traffic aggregates merge or split and which ASs contribute to each flood. For this analysis, only the \textit{outgoing} attack traffic toward the target is considered at each CN node. The \textit{DIST} scenario is configured with moderate traffic volumes, as its goal is to highlight differences in perspective across CN nodes rather than to stress throughput or packet rates.

\subsection{Attack Vector Variability and Composition}

This experiment evaluates the generation of diverse attack vectors with configurable protocol, packet size, data rate, and volume. Scenario \textit{VAR1} combines per-attacker and per-vector settings to produce four distinct pulses. Each pulse maintains an average rate of 5\,Mbit/s but differs in protocol composition, traffic distribution, and fingerprint characteristics (\cf Table~\ref{tab:eval:vec_breakdown}).

\begin{table}[t]
\caption{Attack vector composition}
\label{tab:eval:vec_breakdown}
\centering
\resizebox{\columnwidth}{!}{
    \begin{tabular}{l|cccc}
    \toprule
     & \textbf{V1} & \textbf{V2} & \textbf{V3} & \textbf{V4} \\
    \midrule
    \textbf{Protocol} & TCP SYN & UDP & ICMP & Mixed \\
    \textbf{Packet size [B]} & 42 & 96 & 128 &
    \begin{tabular}[c]{@{}l@{}}36 (49\%)\\48 (18\%)\\96 (6\%)\\128 (10\%)\\256 (17\%)\end{tabular} \\
    \textbf{Avg.\ rate [Mbit/s]} & 4.99 & 5.00 & 5.00 & 4.99 \\
    \textbf{Avg.\ packet rate [pps]} & 14\,874 & 6\,506 & 4\,880 & 7\,065 \\
    \bottomrule
    \end{tabular}
}
\end{table}

\begin{figure}[h]
  \centering
  \begin{subfigure}{0.49\columnwidth}
    \centering
    \includegraphics[width=\linewidth]{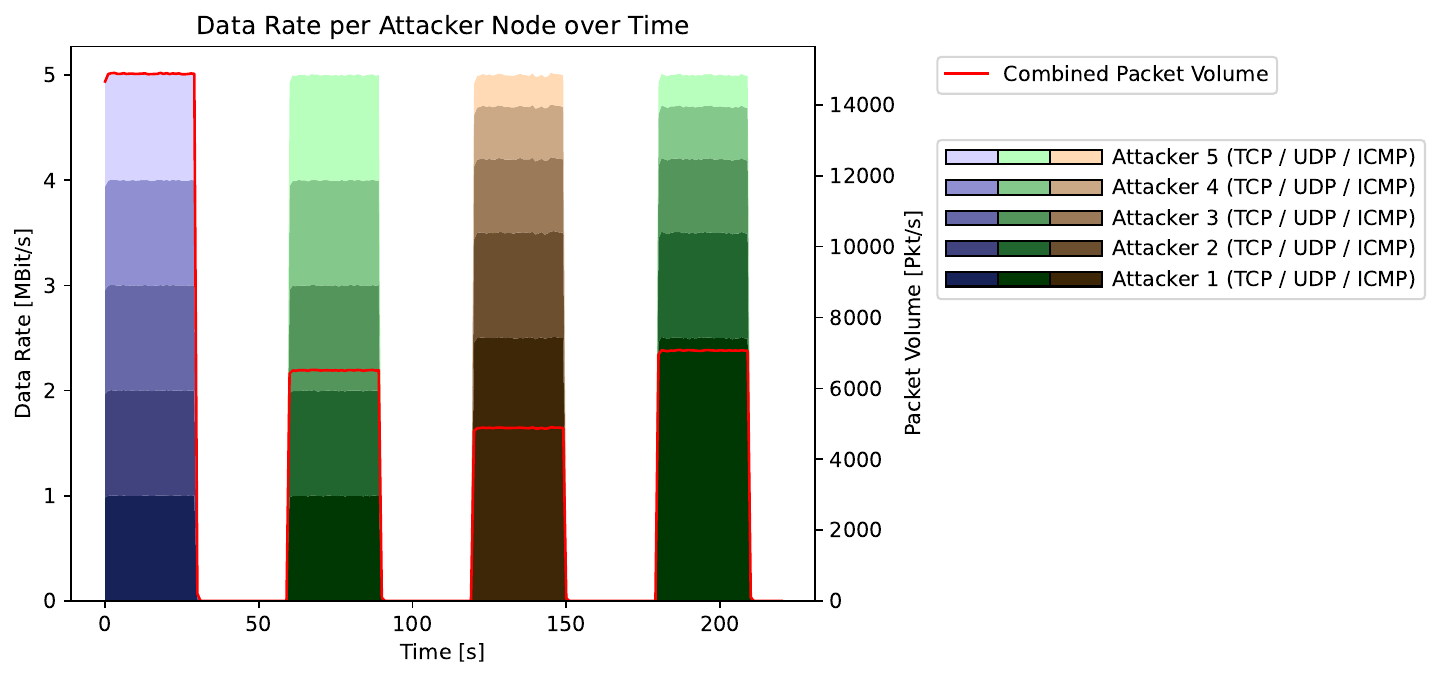}
    \caption{Data rate per attacker and protocol.}
    \label{fig:eval:variability_DR}
  \end{subfigure}%
  \hfill
  \begin{subfigure}{0.49\columnwidth}
    \centering
    \includegraphics[width=\linewidth]{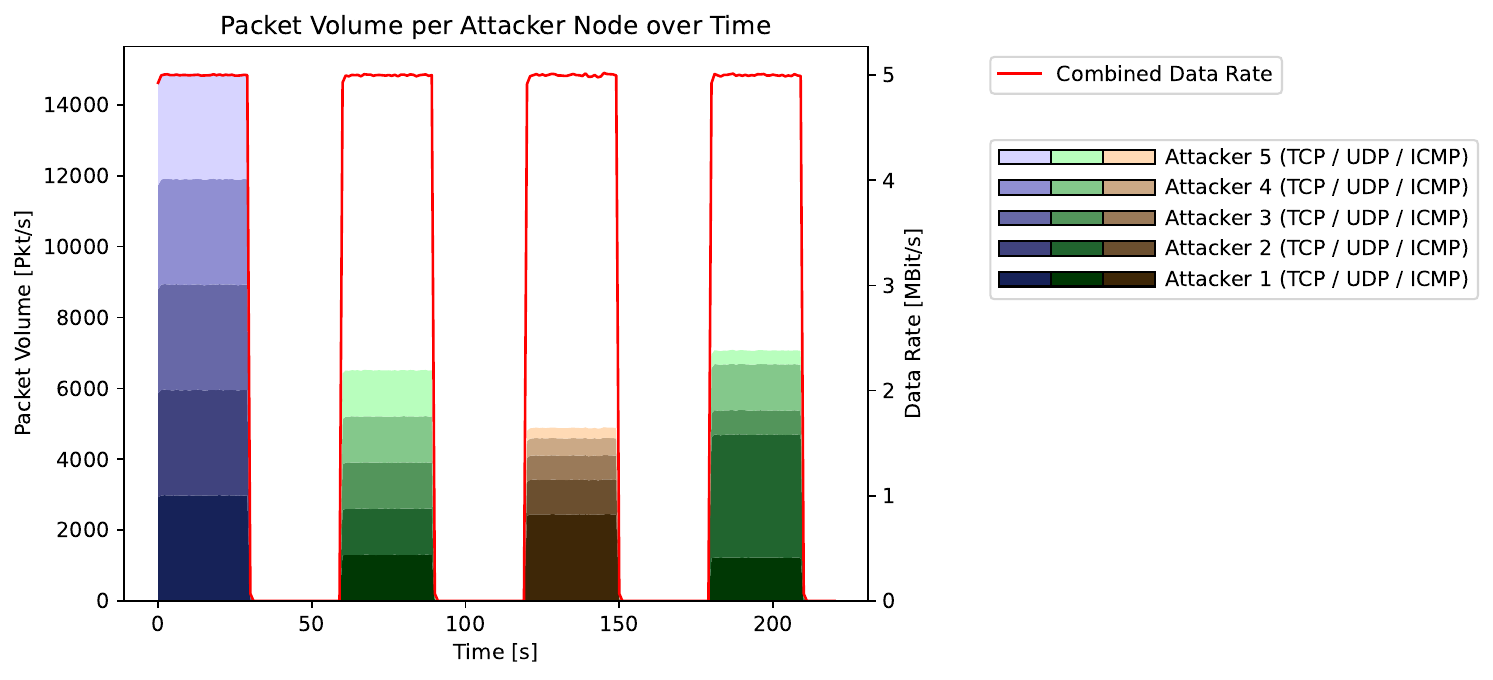}
    \caption{Packet volume per attacker and protocol.}
    \label{fig:eval:variability_PV}
  \end{subfigure}
  \caption{Per-attacker breakdown of traffic composition. Each pulse achieves similar overall rate but distinct packet and protocol characteristics.}
  \label{fig:var1_plots}
\end{figure}

Figure~\ref{fig:var1_plots} shows that identical aggregate throughput can result from distinct packet behaviors. TCP SYN flooding yields the highest packet rate due to its small, fixed-size packets, whereas UDP and ICMP flooding reach the same throughput using larger payloads and fewer packets. The flexibility of configuration allows to assign distinct packet sizes or data rates to individual attackers, resulting in asymmetric contributions that emulate well-orchestrated attacks or irregular floods driven by IoT devices. Source and destination ports can also be fixed or randomized at the vector or node level, introducing additional signature diversity.

\subsection{Variable Pulse-wave Patterns}
\label{ssctn::eval::variability::tiings}

DPWS supports temporal diversity in pulse duration, switching time, and magnitude (data rate and packet volume). Two configurations illustrate this range. \textit{VAR1} uses four homogeneous pulses with identical average data rate and fixed per-pulse parameters. \textit{VAR2} intentionally varies per-pulse data rate, pulse length, and target switch duration to expose the system's temporal flexibility.

\begin{figure}[h]
  \centering
  \begin{subfigure}{0.49\columnwidth}
    \centering
    \includegraphics[width=\linewidth]{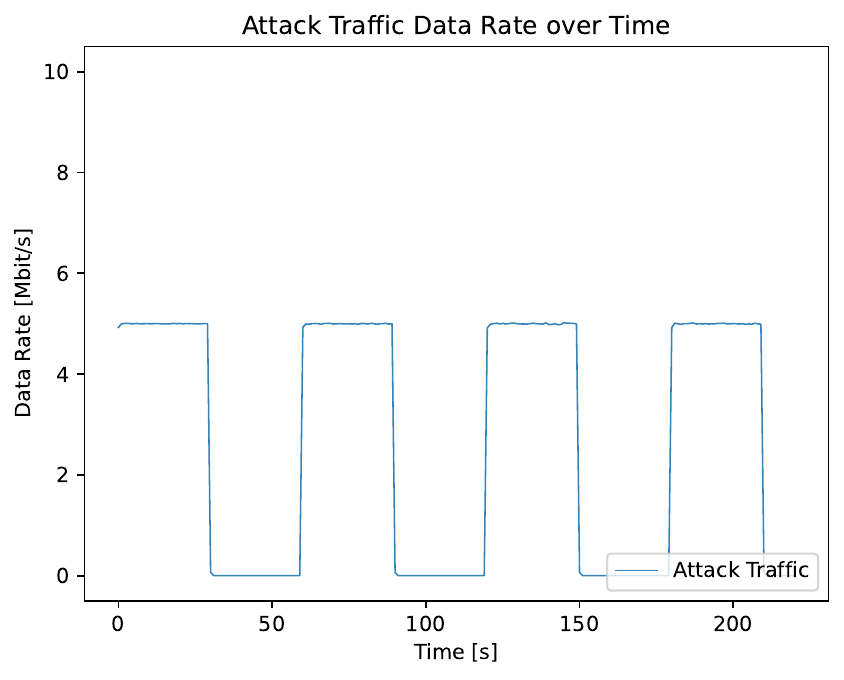}
    \caption{VAR1: constant pulse pattern.}
    \label{fig::eval::timings_var1}
  \end{subfigure}%
  \hfill
  \begin{subfigure}{0.49\columnwidth}
    \centering
    \includegraphics[width=\linewidth]{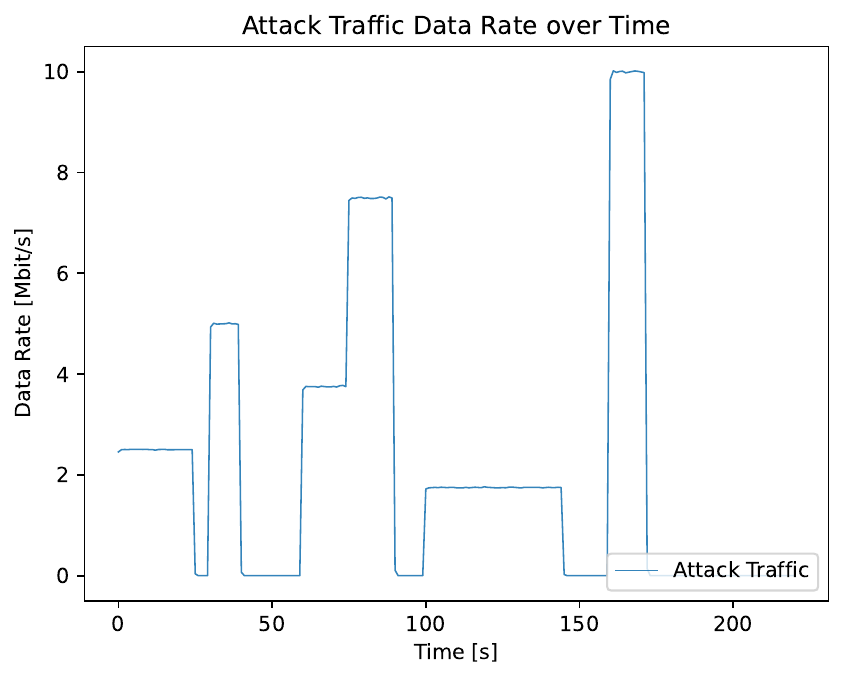}
    \caption{VAR2: variable pulse pattern.}
    \label{fig::eval::timings_var2}
  \end{subfigure}
  \caption{Pulse wave patterns at an IXP capture point. VAR1 uses uniform pulses. VAR2 varies duration, rate, and switch timing.}
  \label{fig:eval:timings}
\end{figure}

Figure~\ref{fig:eval:timings} presents raw traffic time series for both configurations at an IXP capture point. VAR1 shows repeated pulses with consistent shape and amplitude. VAR2 produces heterogeneous pulses with different heights and widths and with both seamless target handovers and explicit switching gaps. A switch duration of zero yields an immediate handover between targets and preserves continuous packet generation, which matches realistic pulse-wave behaviours where attackers redirect capacity without stopping transmission.

\subsection{Dataset Output and Artifacts}
\label{sctn::eval::outputs-artifacts}

DPWS produces synchronized datasets that are suitable for cross-domain analysis. Each experiment exports vantage-specific \textit{.pcap} traces and aggregated flow records together with control-plane logs. These \textit{.pcap} traces preserve temporal alignment across all CN nodes, enabling correlation of data- and control-plane reactions during distributed attacks. Generated datasets can be replayed for offline analysis or used as reproducible inputs for evaluating inter-domain detection and mitigation frameworks.

\subsection{System Scalability}
\label{sctn::eval::scalability}

Scalability was evaluated using three scenarios that scale topology and attack intensity (Table~\ref{tab::eval::scalability_basic_scenarios}). The setup mirrors the SwissIX layout~\cite{design_swissixp} with up to six IXP nodes and pulse-wave attack behaviour. Each run lasted ten minutes and was executed on a 12-core Apple M2 Max with 32\,GB RAM using the optimized \textit{ns-3} build with MPI parallelism. UDP, ICMP, and TCP SYN vectors were configured with equal packet sizes for comparability.

\begin{table}[h]
\caption{Scalability Scenarios}
\label{tab::eval::scalability_basic_scenarios}
\centering
\resizebox{\columnwidth}{!}{
\begin{tabular}{l|ccc}
\toprule
\textbf{Scenario} & \textbf{SC1} & \textbf{SC2} & \textbf{SC3}\\
\midrule
CN nodes & 2 & 4 & 6\\
ASs & 2 & 6 & 12\\
Attackers & 5 & 15 & 30\\
Benign nodes & 10 & 20 & 60\\
Targets & 3 & 3 & 3\\
Non-target servers & 4 & 6 & 12\\
Avg.\ packet rate [pps] & 14\,800 & 66\,900 & 178\,500\\
\bottomrule
\end{tabular}}
\end{table}

Memory leaks observed during early tests were traced to the default MPI synchronization algorithm and eliminated by switching to the null-message strategy, which stabilized RAM use. CPU utilization remained constant at full load, so execution time served as the main scalability indicator.

\begin{table}[t]
    \caption{Execution Time per Scenario}
    \label{tab::eval::scalability_basic_results}
    \centering
    \begin{tabular}{l|ccc}
    \toprule
    \textbf{Scenario} & \textbf{SC1} & \textbf{SC2} & \textbf{SC3}\\
    \midrule
    Avg.\ packet rate [pps] & 14\,800 & 66\,900 & 178\,500\\
    Runtime [s] & 153 & 887 & 3\,821\\
    \bottomrule
    \end{tabular}
\end{table}

Runtime grows faster than packet volume: a 4.5× increase in pps from SC1→SC2 led to a 5.8× longer runtime, and 2.7× from SC2→SC3 to a 4.3× longer runtime. To isolate factors, SC2 variants modified one parameter at a time (Table~\ref{tab::eval::scaling_factors}). Packet volume dominated performance, while changes in benign nodes, IXP nodes, or non-target servers had negligible impact.

\begin{table}[t]
\caption{Impact of Individual Scaling Factors}
\label{tab::eval::scaling_factors}
\centering
\resizebox{\columnwidth}{!}{
    \begin{tabular}{l|l|c}
    \toprule
    \textbf{Variant} & \textbf{Modification vs.\ SC2} & \textbf{Runtime [s]}\\
    \midrule
    SC2 & – & 887\\
    SC2\_AN & 2× attackers, same PPS & 851 ($\downarrow$4.1\%)\\
    SC2\_AS & 2× ASs & 1\,112 ($\uparrow$25.4\%)\\
    SC2\_BN & 2× benign nodes & 859 ($\downarrow$3.2\%)\\
    SC2\_CN & +2 IXP nodes & 909 ($\uparrow$2.5\%)\\
    SC2\_NT & 2× non-target servers & 888 ($\uparrow$0.1\%)\\
    SC2\_PV & PPS as SC3 (2.7×) & 2\,790 ($\uparrow$314\%)\\
    \bottomrule
    \end{tabular}}
\end{table}


\subsection{Lessons Learned}

Developing DPWS in \emph{ns-3} revealed several practical challenges. First, early prototypes using CSMA channels suffered from severe packet loss under high load, which motivated a switch to point-to-point links and per-channel subnets in \textit{ns3} to ensure reliable forwarding. Second, the default MPI synchronization algorithm caused progressive memory leaks, resolved only after adopting the null-message strategy, which stabilized RAM use and enabled multi-core scalability. Third, packet volume (not topology size or benign traffic) proved to be the dominant performance driver. These experiences show that scalable distributed simulations depend more on efficient event scheduling and memory management than on model complexity.

%% file: sections/5_final_considerations.tex
\section{Summary and Future Work} \label{sec:final_considerations}

This work introduced DPWS (source-code \cite{dpws-code}), a simulator for distributed pulse-wave DDoS datasets with synchronized multi-vantage visibility, reflecting experiences with large-scale ns-3 synchronization and stability. Experiments confirmed realistic pulse dynamics and variability, showing how identical rates can conceal distinct packet behaviors and underscoring the need for distributed correlation. Future work will extend dataset realism through adaptive traffic shaping and integration of inter-domain signaling for coordinated mitigation testing.